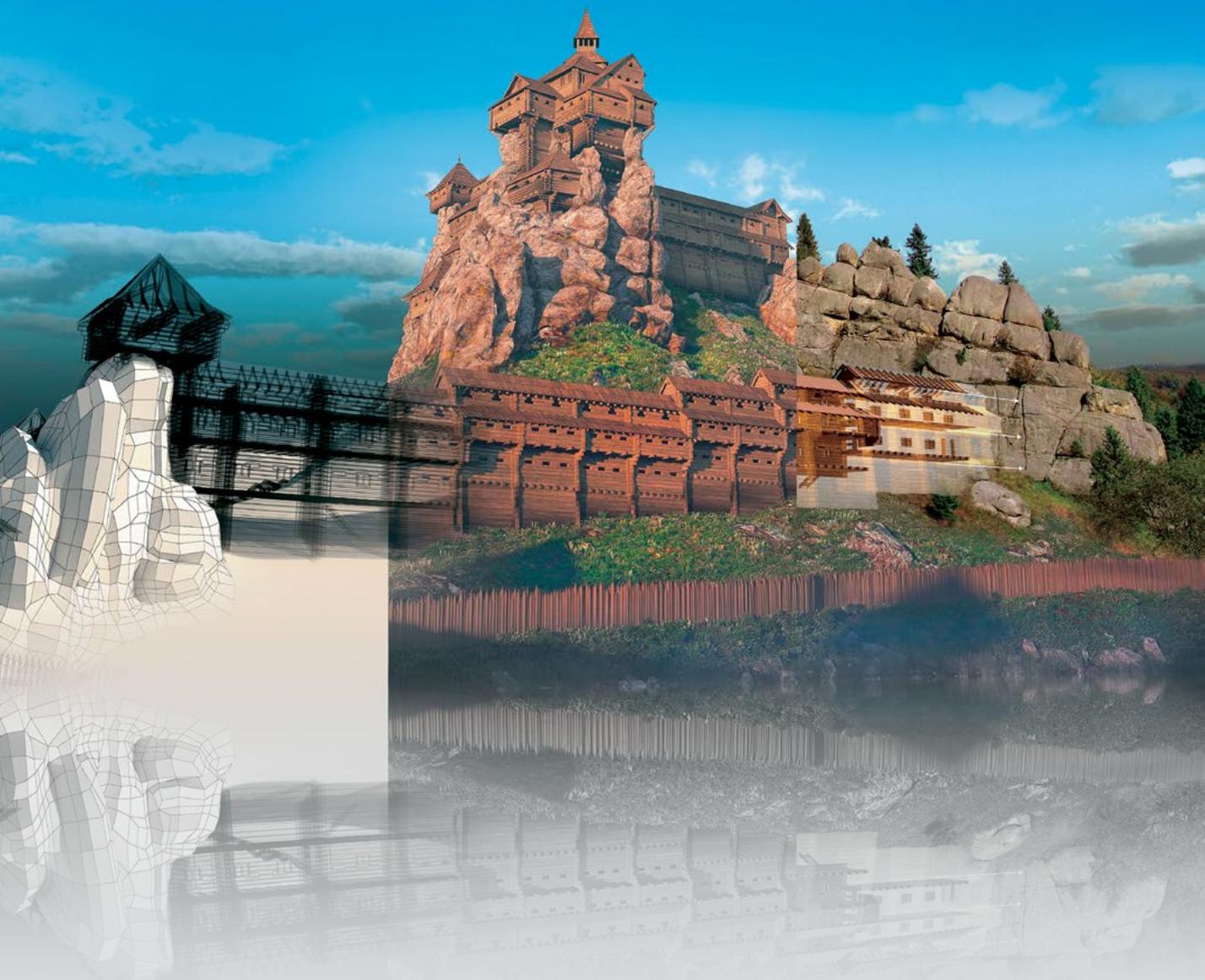

Virtual reconstruction of Tustan Fortress.

Olga
Barkova,
Natalia
Pysarevska,
Oleg
Alienin,
Serhii
Hamotsky,
Nikita
Gordienko,
Vladyslav
Sarnatskyi,
Vadym
Ovcharenko,
Mariia
Tkachenko,
Yurii
Gordienko,
Sergii
Stirenko
Ukraine

# 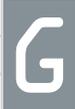amification for Education of the Digitally Native Generation by Means of Virtual Reality, Augmented Reality, Machine Learning, and Brain-Computing Interfaces in Museums

## Abstract

Particularly close attention is being paid today among researchers in social science disciplines to aspects of learning in the digital age, especially for the Digitally Native Generation. In the context of museums, the question is: how can rich learning experiences be provided for increasingly technologically advanced young visitors in museums? Which high-tech platforms and solutions do museums need to focus on? At the same time, the software games business is growing fast and now finding its way into non-entertainment contexts, helping to deliver substantial benefits, particularly in education, training, research, and health. This article outlines some aspects facing Digitally Native learners in museums through an analysis of several radically new key technologies: Interactivity, Wearables, Virtual Reality, and Augmented Reality. Special attention is paid to use cases for application of games-based scenarios via these technologies in non-leisure contexts—and specifically for educational purposes in museums.

## Introduction

Over the past 50 years, interest in education and science has undergone several ups and downs in the European Union and worldwide, especially among youth. It had pronounced peaks at the start of the atomic era in 1945, the first launch of a Soviet space satellite in 1957, the exploration of the ocean depths by Cousteau's team in the 1960s, the invention of the personal computer in the early 1980s, and so on. But the high peaks were frequently followed by yawning abysses—for example, after the end of the American Apollo program for the exploration of the moon in the early 1970s, the shutdown of the Superconducting Super Collider in the USA in 1993, the dot-com crash and blow-up of the fictitious "Millennium Bug" in 2000, and so on. The modern trend of growing youth interest in the glamourous lifestyles of pop music, fashion, and movie "stars" is not only a result of successful advertising and public relations in the entertainment business, but also an inability of the current educational system to compete with the globalized allure of "show business". As a result, the number of young people wanting to become students of natural sciences has dropped significantly (Lester, 2007).

For more than a decade, the European Union (EU) has carried out regular surveys in a program called Eurobarometer (European Commission, 2006), to measure public opinion and knowledge on a variety of topics across its member states. One reason is to find common ground as the EU makes policies for countries with diverse cultures; another is to evaluate the effects of past EU programmes. The goal of the Eurobarometer on "Europeans, Science and Technology (S&T)" was to determine the interest and level of informedness of European citizens; their image and knowledge of S&T; their attitudes towards S&T; their ideas about the responsibilities of scientists and policy-makers; and their perception of scientific research in Europe compared with other parts of the world. On the whole, there was a noticeable drop in the number of people who claimed to be "very interested" in scientific topics between 1992 and 2005. The lack of relevance of the S&T curriculum is probably one of the greatest barriers for successful learning as well as for interest in the subject.

This is especially unfortunate in view of the recent great advances of modern science, especially on the international scale: the success of unprecedented international cooperation efforts dedicated to the construction of the Large Hadron Collider in CERN (Allen, 2016); the unprecedented simultaneous investigations of four objects (Moon, Venus, Mars, and Saturn) in our solar system by a single agency, the European Space Agency (ESA) (2017); the European Grid Initiative (EGI) (Kranzlmüller, et al, 2010) ; beginning construction of the International Thermonuclear Experimental Reactor (ITER) (Bigot, 2015); and so forth.

The main aim of this paper is to review some radically new technologies (Interactivity, Virtual Reality, Augmented Reality, Wearables, Machine Learning, and Brain-Computer Interface) that are currently used by authors in games-based scenarios in non-leisure contexts–and specifically for educational purposes in museums. The current state of the art and the available approaches are outlined briefly in Section 2 – Problem and Related Work. Some radically new technologies, like Virtual Reality, Augmented Reality, Machine Learning, Multimodal Interactivity, Wearables, and Brain-Computer Interface, with the gamification use cases under investigation are presented in Section 3 – Information Technologies for Gamification. Then the various aspects of these techniques and conclusions as to their implementation in gamification context are discussed in Section 4.

## Problem and Related Work

Modern society simultaneously exhibits the entire spectrum of functional / computer / digital literacy, including absolute computer illiteracy (mostly the elderly), digital phobic, basic computer literacy, digital immigrants, intermediate computer literacy, digital visitors, proficient computer literacy, digital residents, and digitally native (Prensky, 2001;White, Le Cornu, 2011). Usually youth (in primary and secondary schools and higher education, including university level) are very used to technology in their play and everyday life: just think of current videogame consoles that offer hundreds of hours of play, with no intellectual growth at all. They are users of electronic products made by science, but the question "how does it work?" is rarer and rarer. They lack curiosity – or worse, they cannot find teachers to answer their questions (if any). This clamorous contradiction highlights the importance of attracting young people toward science careers from the earliest ages: not only from the universities, but also among pupils of secondary and even primary schools.

In the context of museums, the question is: how can rich learning experiences be provided for increasingly technologically advanced young visitors in museums? Which high-tech platforms and solutions do museums need to focus on? At the same time, the software games business is growing fast and now finding its way into non-entertainment contexts, helping to deliver substantial benefits, particularly in education, training, research, and health. In this way, museums can be the place where games and game-like scenarios of interaction can be proposed to the Digitally Native Generation with an aim of increasing interest in education and science (Martí-Parreño et al, 2016). Below we consider some aspects facing Digitally Native learners in museums through an analysis of several radically new key technologies: Interactivity, Wearables, Virtual Reality, and Augmented Reality. Special attention is paid to use cases for application of games-based scenarios via these technologies in non-leisure contexts—and specifically for educational purposes in museums.

## Information Technologies for Gamification

The fast development of some radically new technologies, like Indoor Geolocation, Virtual Reality, Augmented Reality, Machine Learning, Multimodal Interactivity, Wearables, and Brain-Computing Interfaces, etc., allows us to use the associated hardware and software for a wide range of users, including ordinary visitors of museums, and especially the youngest.

### Indoor Geolocation

Geolocation and positioning technologies are the most important aspects of developing applications for mobile devices. In addition to navigation, more and more often, location data are used for entertainment, cognitive, and promotional purposes. In this regard, mobile devices are necessary for the most accurate determination of the coordinates, which the well-known and available methods do not always cope with. This article provides an overview of existing solutions for increasing the accuracy of positioning mobile devices using Wi-Fi access points, Bluetooth beacons, and location data from the Google API (Martí-Parreño et al, 2016).

Consider ways to implement an application for museums, exhibitions, architectural and other attractions. The application should accurately determine the position of the device near an object and provide the user with information about it in the form of text, audio, or video files.

Let us now consider examples of applications that solve the problem of popularizing museums, exhibits, and any other memorable places through the Internet of Things (IoT). This IoT technology allows you to interact with physical objects with the help of computing devices that everyone owns (mobile devices in this context). A database of objects-exhibits is needed to implement this idea. It will store any kind of information—for example, simple text information, video materials, and 3D models of exhibits for visualization with augmented reality (AR).

There are two options for implementing a mobile application:

- Local—for each museum a separate application is created with its own database of exhibits;

- Global—a single database of exhibits is created and a single application that covers a larger number of users, which provides many options for increasing museum attendance, monetization, etc.

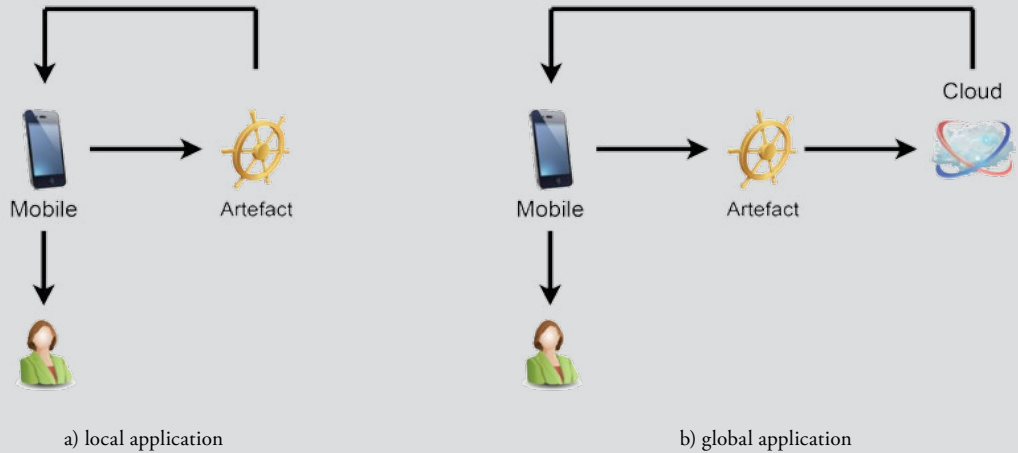

a) local application                                  b) global application

Fig. 1. Simple diagram of the local and global application

The global application is of great interest because of its portability to various museums. The user will be able to open the application and see a map with the labels of museums, memorial places, and objects. General information about the object will be opened by clicking on the label, with the ability to view the list of exhibits and, the schedule of work, call button, to get directions, and, perhaps, to purchase tickets with one click in the future. Also, the user can search or filter information to find a suitable museum or a place to visit.

When the user is on the museum grounds' territory next to the exhibit, he sees information about the nearest exhibit in the form of descriptions, photographs, pictures, stories, video materials, etc. In the case of QR codes, the user must find this label, select the "Scan" menu item, and then capture the QR code using the camera, after which he will obtain information about the exhibit. For museums with large items (like the Museum of Folk

Architecture and Folkways of Ukraine in Pyrohiv, in the suburbs of Kiev, Ukraine) QR codes can be replaced with GPS / WiFi positioning, but it is desirable to combine these two methods when the user does not want to include GPS.

### Augmented Reality

As is well-known, augmented reality (AR) provides a direct view on a real world and an indirect view on some added (augmented) computer-generated elements, such as sound, graphics, video, or other data. In the context of museums, a place or territory where it is planned to use this AR-related technology should be "labelled" ("AR-geotags" should be attached) so that visitors can easily find the places of interest. The visitors follow the AR-geotags and see the AR-marker. Then they can point the smartphone camera at the

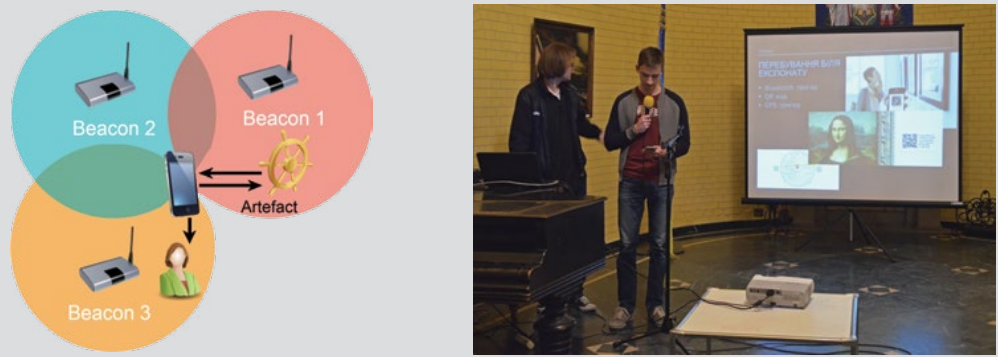

Fig. 2. Indoor geolocation in a museum: triangulation principle of indoor geolocation (left), and usage of smartphone as a context-sensitive guide with the more detailed information shown on the screen in the National Historic and Architectural Museum "Kiev Fortress" (https://www.kyiv-fortress.org.ua/)

AR-marker and show additional relevant information—"geolayer" as in Fig. 3 (left), and / or augmented reality with multimedia information (3D models, text, video, audio, etc.) as in Fig. 3 (right). Moreover, visitors can add their own AR-geotags (or comments to existing tags) and content to such an open "digital ecosphere" driven by AR and other volunteers.

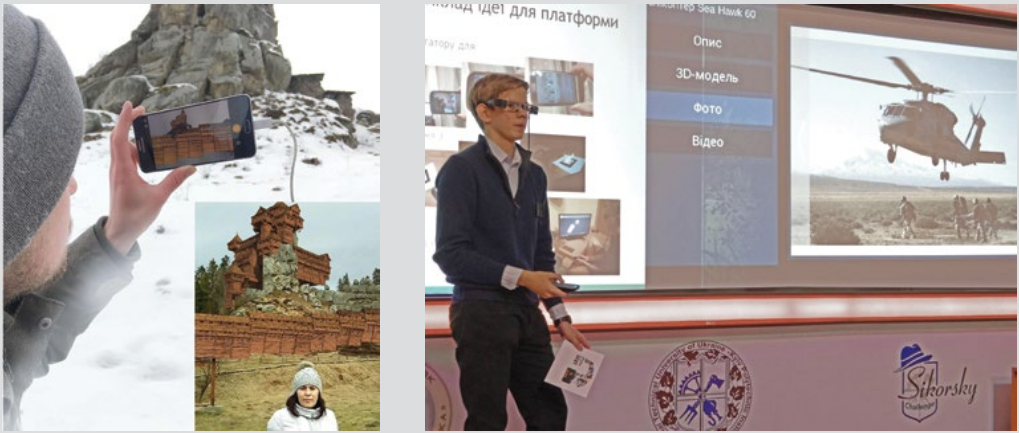

Fig. 3. The brilliant example of the AR application on the landscape of the missing historical object is the Tustan AR addition for visualizing the missing fortress (Vasyl Rozhko, *Tustan Fortress, Ukraine: Virtual Reality of the Past, a Cultural Landscape for the Future*, (http://network.icom.museum/fileadmin/user_upload/minisites/icamt/ICAMTYEARS/2011_-_2020/2016/2016_Milan/2016_Milan_3_papers/20_Vasyl_Rozhko.pdf, left), and contextual multimedia information dedicated to Igor Sikorsky in the State Polytechnic Museum at Igor Sikorsky Kyiv Polytechnic Institute (http://museum.kpi.ua, right)

### Virtual Reality

In contrast to augmented reality (which enhances perception of reality), virtual reality (VR) replaces the real world with a simulated one by VR-headsets, sometimes in combination with other components, to generate realistic images, sounds, and other sensory inputs that imitate a physical presence in some imaginary environment. Visitors in museums can use VR-equipment (like a smartphone inside Google Cardboard in Fig. 4) to "look around" the artificial world and interact with virtual components inside it. This technology is not very new in the museum context. Even the British Museum and the Guggenheim, for example, convert some of their content to a virtual reality representation and deliver it to their visitors.

### Machine Learning for Note Recognition from Authentic Music Instruments

Conversion of audio files into musical notation (music transcription) is a popular and very difficult problem even for musicians and experts. That is why the available music transcription tools and methods hardly compete with human perception (Gowrishankar, Bhajantri, 2016). Recently, several solutions for audio search (Shazam, Soundhound, Doreso) have been proposed. For example, in 2003, Avery Li-Chun Wang, a chief scientific engineer at Shazam, introduced the audio search algorithm (Wang, 2003), where a microphone was used to pick up an audio sample. Then a digital summary of the sound was generated as an acoustic fingerprint – i.e. it was broken down into a simple numeric signature that was unique to each track and then matched to an extensive audio music database. These algorithms are known to perform well on recognizing a short audio sample of music that had been broadcast, mixed with heavy ambient noise, subject to reverb and other processing, captured by a poor microphone, subjected to voice codec compression, etc. Today, rethinking the search-by-sound problem within the context of current machine learning advances could produce surprising results and possibly reveal some intricacies of human hearing that are still not understood (Sarnatskyi et al, 2017).

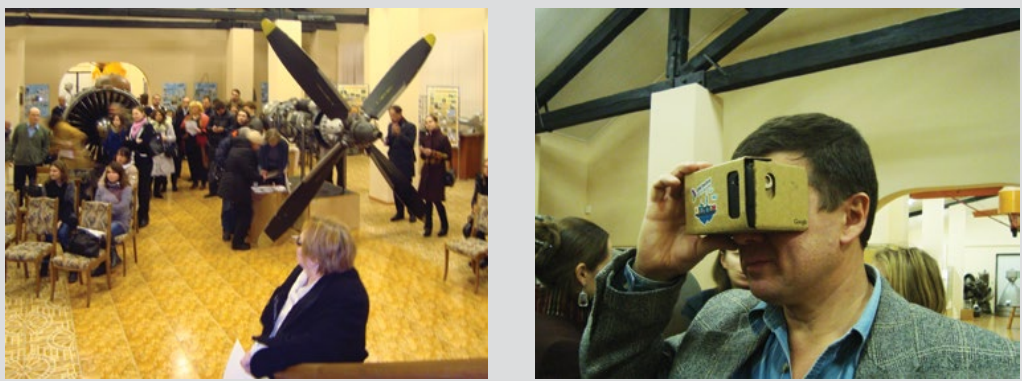

Fig. 4. Virtual reality in the museum: real exposition in the State Polytechnic Museum at Igor Sikorsky Kiev Polytechnic Institute (http://museum.kpi.ua) (left); virtual representation of some objects by means of Google Cardboard for VR (right)

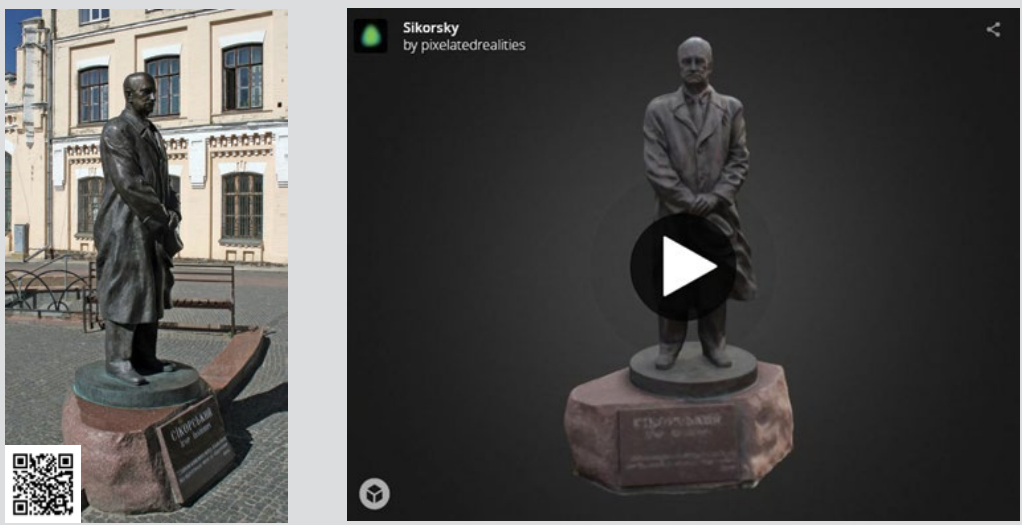

Fig. 5. Igor Sikorsky monument at Igor Sikorsky Kyiv Polytechnic Institute with QR-code with web-link for access to the web-page of the object with the supporting multimedia material (left), including the virtual representation of the monument (right) by Pixelated Realities (http://pixelatedrealities.org)

The combination of machine learning, digital signal processing, and mobile computing allows us to create mobile applications that can analyse completely unknown music and create the corresponding sequence of music notes (Fig. 6).

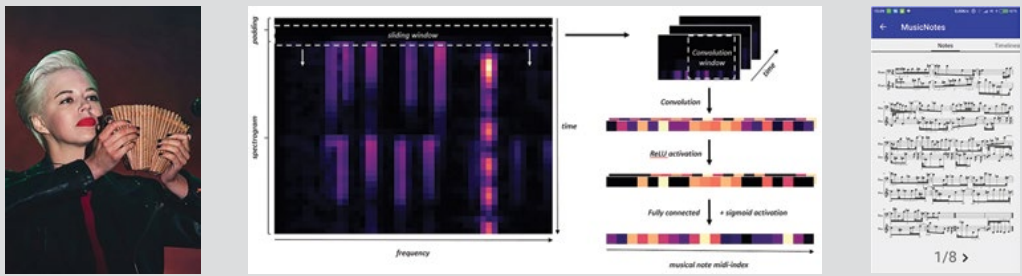

Fig. 6. Typical example of machine learning used for music analysis in the museum: recording an ancient authentic musical instrument as used by the Onuka electric folk band (photo by Andriy Marchenko, public domain) (left), constructing spectrogram and machine learning (center), and final note recognition and transcription (right)

## Machine Learning for Historic Symbol Decoding

The necessity to create and use writing systems has been with humankind since the dawn of time, and they have always evolved based on the concrete challenges the writers and readers faced. For example, the angular shapes of runes are very convenient to be carved in wood or stone, but they are not evident for modern readers without expert knowledge of them (Williams, 1996). Current progress in machine learning allows us to train a neural network to recognize ancient writing and decode it for the general public on the basis of the methodology for decoding the handwritten symbols like digits (Kochura et al, 2017) and shorthand (Hamotskyi et al, 2017). Of course, it takes special training for the different cases of such symbols like the runes (Fig. 7c) (Williams, 1996) or the authentic graffiti on the walls of St. Sophia Cathedral (Kiev, Ukraine) (Fig. 7d) (Kornienko, 2011).

### Brain-Computing Interface and Multimodal Interaction

Brain-Computer Interfaces (BCIs) are widely used to study interactions between brain activity and the environment. This research is often oriented toward mapping, assisting, augmenting, or repairing human cognitive or sensory-motor functions. The most popular signal acquisition technology in BCI is based on measurements of EEG activity. It is characterized by different wave patterns in the frequency domains or "EEG rhythms": Alpha (8 Hz–13 Hz), SMR (13 Hz–15 Hz), Beta (16 Hz–31 Hz), Theta (4 Hz–7 Hz), and Gamma (25 Hz–100 Hz). They are related with various sensorimotor and/or cognitive states, and translating cognitive states or motor intentions from different rhythms is a complex process, because it is hard to associate directly these frequency ranges to some brain functions. Some consumer EEG solutions, such as MindWave Mobile by Neurosky, Muse by InteraXon, Emotiv EPOC by Emotiv and open source solutions like OpenBCI have become available recently and can be used to assess emotional reactions, etc. (Stirenko et al, 2017).

Through EEG measurements they can determine different psychophysiological states, such as attention, relaxation, frustration, or others. For example, MindWave Mobile by Neurosky can determine at least two psychological states: concentration ("attention") and relaxation ("meditation"). The exposure of a user to different external stimuli will change both levels of the psychological states collected by this device. For example, if the user is calm, then relaxation ("meditation") will be high and concentration ("attention") will be low. The consumer BCI-devices have different numbers of EEG channels, additional sensors, and types of EEG connections with human surfaces. Their prices depend on their possibilities.

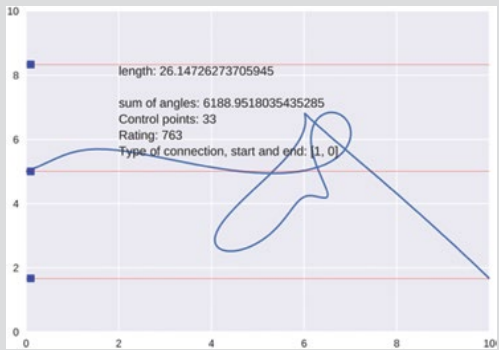

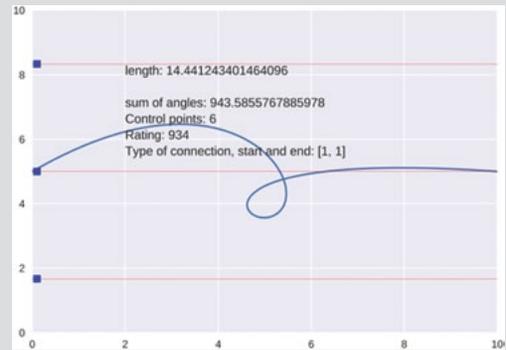

a)                                b)

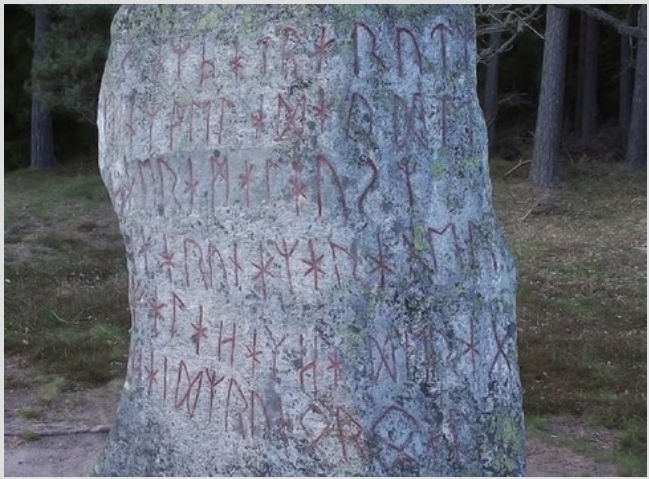

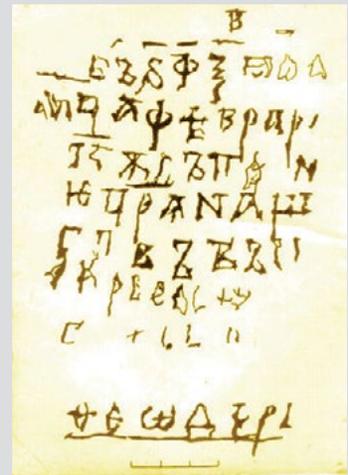

c)                                d)

Fig. 7. Example of generated glyphs with low (a) and high (b) fitness used for shorthand generation and estimation; and the 6-7th-century runic inscription (Björketorp Runestone, Blekinge, Sweden, photo by Henrik Sendelbach, public domain) (c); the ancient 10th-century graffiti about the death of the Grand Prince of Kievan Rus, Yaroslav the Wise, on the walls of Saint Sophia's Cathedral (https://st-sophia.org.ua/en/home/) (d)

This general concept of multimodal integration was verified by an experimental setup (Fig. 8). It includes smart glasses Moverio BT-200 by EPSON as a visual AR interaction channel for the controlled cognitive load (learn about the set of historical artefacts in the museum) and a collector of accelerometer data; neurointerface MindWave by NeuroSky as a BCI-channel and collector of EEG-data; and heart monitor UA39 by Under Armour as a collector of heartbeat data (Gordienko et al, 2017). The setup can collect time series of several parameters: subtle head accelerations (like tremors characterizing stress), EEG-activity, and intervals of heartbeats on the scale of milliseconds. Statistical methods were used to find correlations between these time series for various conditions, and machine learning methods were used to determine and classify various regimes.

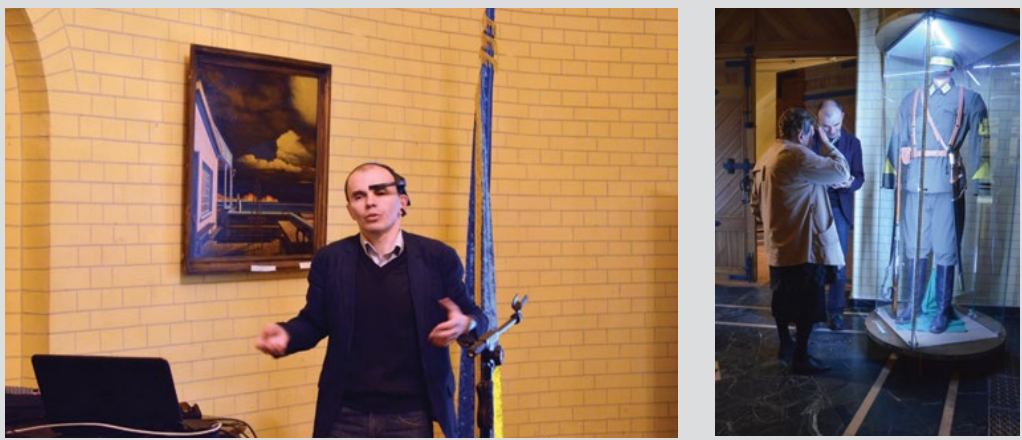

Fig. 8. Brain-computing interface in a museum: estimate the level of concentration/relaxation among guides and visitors to get feedback information on the attractiveness of exposition and cognitive load on visitors in the National Historic and Architectural Museum "Kiev Fortress" (https://www.kyiv-fortress.org.ua/)

The previous experiments with the general concept and setup allowed us to propose several possible applications for the integration of AR channels with BCI technologies to provide direct neurophysical feedback on the attractiveness of exposition and cognitive load on visitors. The general idea of AR-BCI integration is based on the establishment of multimodal interactions and data flows, where all EEG reactions from the user observed by various available BCI-ready devices or combined AR-BCI devices are gathered and then processed by machine learning methods on the supporting devices (smartphone, tablet, etc.) in a non-obtrusive way. The essence of the AR-BCI integration consists in the real-time return of the obtained output data of a neurophysical nature to users as AR feedback by available AR-ready gadgets (through sound, visual, and tactile AR channels) (Gordienko et al, 2017).

## Conclusions

Despite the fact that museums have always been considered as educational institutions and used various technologies, museum visitors are ahead of them: they are better connected and informed, they are more mobile and like to get information wherever and whenever they are, rather than passively receive it as it arrives. In this sense, "gamification" through any of the newly available information and communication technologies (ICT) can provide a playful experience that increases motivation, involvement, and fun among visitors. In this paper, several of our approaches for gamification of the learning process through the newly available ICT technologies are presented. All of them were successfully demonstrated during scientific and technical seminars "Digitized Heritage: Preservation, Access, Representation" (https://goo.gl/Wgs5SS) that have taken place in State Polytechnic Museum at Igor Sikorsky Kyiv Polytechnic Institute for five years, during which digital access technologies were demonstrated. The target audience included students, teachers, schoolchildren, families, local history researchers, writers and journalists, artists, general public citizens, and even tourists. As a result, we have summarised our experience on how to propose and use gamification in museums in the following aspects: variety, fun, autonomy, feedback, creativity, and tolerance for creative experiments (even when they are clearly wrong) and "smart creatives" in the sense of the Google corporate culture (Schmidt, Rosenberg, 2014).

The secret ingredient that makes gamification a truly special experience is fun. Fun is a consequence of brain adaptation to pattern recognition—that is, it is a consequence of learning. The traditional belief is that fun only promotes learning, but fun actually plays an essential role in learning. The existence of fun during the tasks of a course is an important indication that learning is occurring, and at the same time the cycle is fed back (due to dopamine), so that our students want to continue to perform more tasks. The results of our activities can be used also to conduct active intellectual leisure such as varied types of excursions, quests and other public events, participating in creating of new content and art. It could be especially effective according to policy recommendations and action points identified for the (re-)use of European digital cultural heritage, collated under Europeana, in education and learning (https://goo.gl/iKEtES).

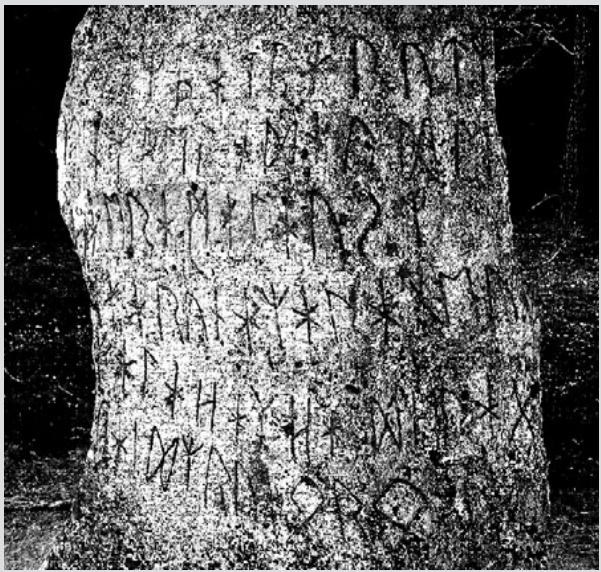